\documentclass[showpacs,aps,prl,preprint,reprint,superscriptaddress,nofootinbib,longbibliography]{revtex4-2}

\usepackage[dvipdfmx]{graphicx}
\usepackage[colorlinks=true, pdfstartview=FitV, linkcolor=magenta,citecolor=blue, urlcolor=magenta,
bookmarks=true, bookmarksnumbered=true, breaklinks]{hyperref}
\usepackage{amsmath,amssymb,bm,color,longtable,mathrsfs,amsfonts,slashed,ulem}
\usepackage[T1]{fontenc}       
\usepackage[utf8]{inputenc}    

\allowdisplaybreaks

\newcommand \beq{\begin{eqnarray}}
\newcommand \eeq{\end{eqnarray}}

\newcommand{\lqcd}{\Lambda_{\rm QCD}}

\newcommand{\br}{ {\bm r} }

\newcommand{\la}{\langle}
\newcommand{\ra}{\rangle}

\newcommand{\calK}{\mathcal{K}}

\newcommand{\rmd}{\mathrm{d}}
\newcommand{\rmi}{\mathrm{i}}
\newcommand{\rme}{\mathrm{e}}

\newcommand{\up}{\uparrow}
\newcommand{\down}{\downarrow}

\begin{document}


\begin{flushright}
\end{flushright}

\title{ Meson molecules in strong magnetic fields:\\
non-monotonic evolution of the charged pion and kaon energies
}

\author{Toru Kojo}
\email{ torukojo@post.kek.jp }
\affiliation{ Theory Center, IPNS, High Energy Accelerator Research Organization (KEK), 1-1 Oho, Tsukuba, Ibaraki, 305-0801, Japan }
\affiliation{ Graduate Institute for Advanced Studies, SOKENDAI, 1-1 Oho, Tsukuba, Ibaraki, 305-0801, Japan }
%

\date{\today}

\begin{abstract}
In strong magnetic fields, charged quarks occupy the lowest Landau level, leading to an effective dimensional reduction of hadronic dynamics.
This dimensional reduction naturally generates a hierarchy of scales,
separating fast intra-meson quark dynamics from slow collective meson motion; this motivates a Born-Oppenheimer description of meson-meson systems.
Our Born-Oppenheimer analysis shows that the infrared behavior is controlled 
by the interplay between dimensional reduction and the structure of the meson-meson interaction, leading to three distinct regimes: scattering-dominated, molecular, and compact multiquark states.
Charged pseudoscalar mesons such as $\pi_+$ and $K_+$ provide a particularly interesting realization of this framework, 
as their lattice spectra at large magnetic fields suggest the emergence of loosely bound states near the boundary between scattering and molecular regimes.
Our results suggest that strong magnetic fields provide a useful laboratory for exploring the emergence and classification of hadronic bound states.

\end{abstract}


\maketitle

\section{Introduction}

\begin{figure}[b]
\vspace{-.4cm}
\begin{center}
\includegraphics[width=8.5 cm]{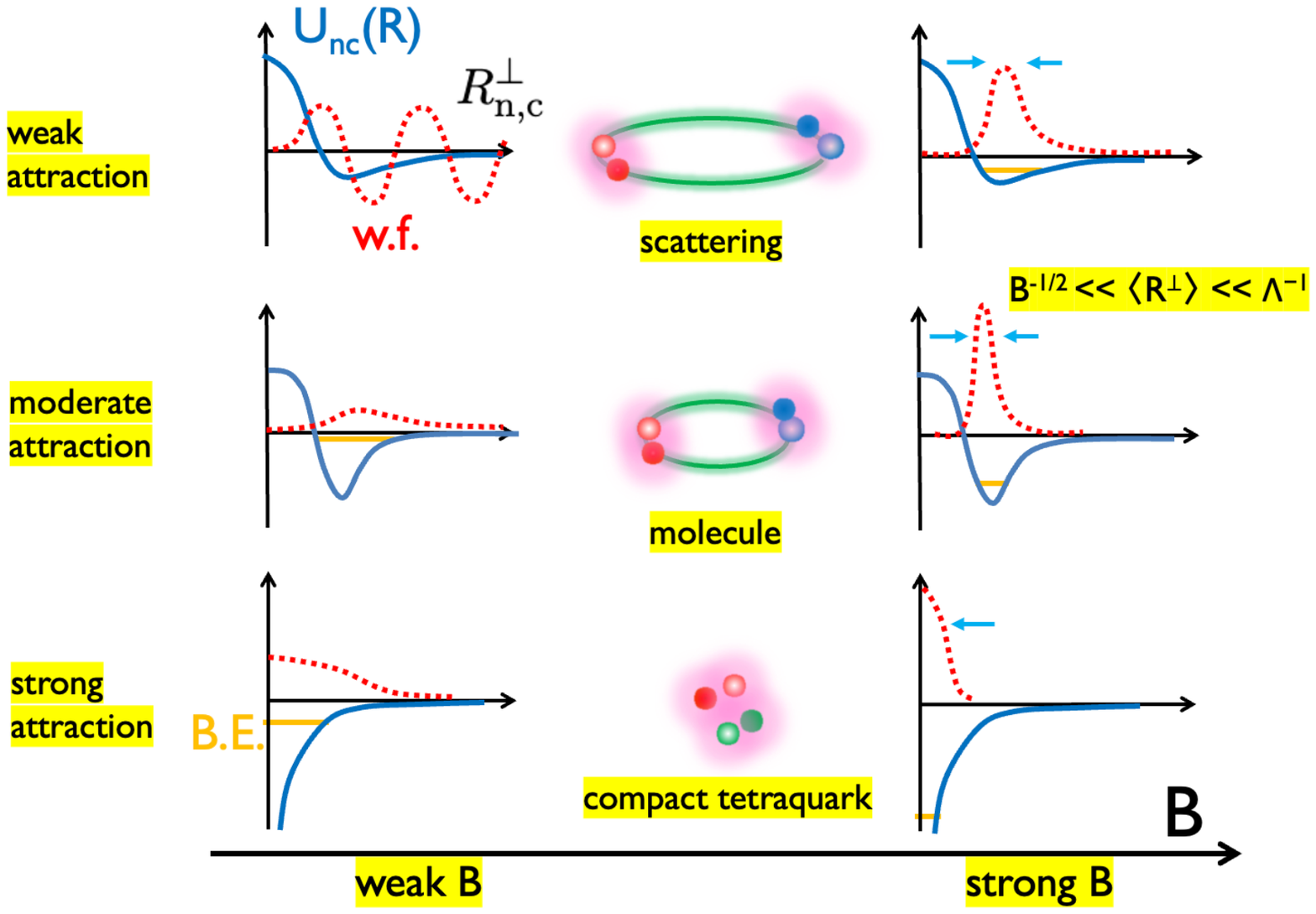}
\end{center}
\vspace{-0.5cm}
\caption{Schematic picture of two-meson dynamics in strong magnetic fields.
Increasing $B$ suppresses transverse quantum fluctuations and reduces the slow collective motion to an effective one-dimensional problem.
Depending on the structure and strength of the meson-meson interaction ($U_{\rm nc}$), 
the system evolves from a scattering-dominated regime to a molecular regime, or to a compact multiquark regime when short-distance attraction dominates.
}
\label{fig:molecule_evolution}
\end{figure}   

Hadronic molecules and compact multiquark states have attracted considerable attention 
as they broaden our view of hadrons beyond the conventional $q \bar{q}$ and $qqq$ pictures \cite{Guo:2017jvc}. 
A central difficulty in identifying such structures is that
molecular, scattering, and compact configurations can mix strongly within the same quantum channel \cite{Esposito:2016noz}. 
This difficulty is particularly severe in light-quark systems, 
where large kinetic energies compete with interactions, 
making simple semiclassical intuition less reliable than in heavy-quark systems.
Indeed, much of the recent progress in identifying exotic hadrons has been achieved in the heavy-quark sector \cite{Lebed:2016hpi}.

Strong magnetic fields offer a distinct control parameter for reorganizing this problem.
By quenching transverse kinetic motion, they effectively reduce hadronic dynamics toward lower dimensions, 
thereby enhancing the role of residual hadron-hadron interactions in determining the infrared structure of hadronic states.
By continuously varying the magnetic field strength, 
one can track the evolution of hadronic states from scattering-dominated configurations toward molecular or compact multiquark regimes.
Although producing such strong magnetic fields in laboratory experiments remains challenging \cite{Huang:2015oca},
current lattice Monte-Carlo simulations can study QCD dynamics in magnetic fields far beyond the QCD scale, $\lqcd \simeq$ 200--300 MeV \cite{Bali:2017ian,DElia:2021tfb}.

In the lowest Landau level (LLL) for quarks, 
transverse motion is quenched and the remaining dynamics becomes effectively one-dimensional, up to model-dependent self-energy corrections \cite{Andersen:2014xxa,Miransky:2015ava}.
When $B$ is applied along with the $+z$ direction, $u_\up, d_\down, s_\down$, and their antiparticles can stay in the LLL,
while the others are in higher Landau levels.
The dimensional reduction also takes place for mesons composed of quarks in the LLL \cite{Fukushima:2012kc}.
For neutral mesons, the transverse kinetic energies are quenched.
Likewise, for a charged meson, 
the energy weakly depends on the growth of $\ell_z$, leaving approximate Landau degeneracy \cite{Kojo:2021gvm}.

In this paper, we show that dimensional reduction in two-meson systems 
generates a hierarchy of scales that allows us to separate fast intra-meson quark dynamics from slow collective meson motion.
Applying a Born-Oppenheimer description of meson-meson systems,
we find that the interplay between the quenched transverse dynamics and meson-meson potentials
leads to three distinct regimes: scattering-dominated, molecular, and compact multiquark regimes.
The resulting physical picture is summarized in Fig.~\ref{fig:molecule_evolution}.

Charged pseudoscalar mesons such as $\pi_+$ and $K_+$ provide concrete realizations of this Born-Oppenheimer framework.
In full lattice QCD \cite{Ding:2020hxw,Ding:2026qzu}, their energies show a puzzling non-monotonic evolution with increasing $B$,
in contrast to the monotonic behavior expected from a simple $q\bar q$ picture,
where the relevant charged pseudoscalar configuration involves a higher Landau level~\cite{Taya:2014nha,Kojo:2012js}.
We argue that $q \bar{q} q \bar{q}$ components beyond the standard $q\bar{q}$ basis provide a natural way to interpret this behavior. 
In quenched simulations, such non-monotonic behavior is absent, consistent with the suppression of $q\bar q \leftrightarrow q \bar{q} q \bar{q}$ mixing \cite{Bali:2011qj}.
While meson spectra in strong magnetic fields have been extensively studied,
mesonic molecular formation induced by magnetic-field-driven dimensional reduction
has not been systematically explored.
Related interpretations based on meson mixing have also been discussed \cite{Wang:2026xsm}.

\begin{figure}[t]
\vspace{-.cm}
\begin{center}
\includegraphics[width=8.5 cm]{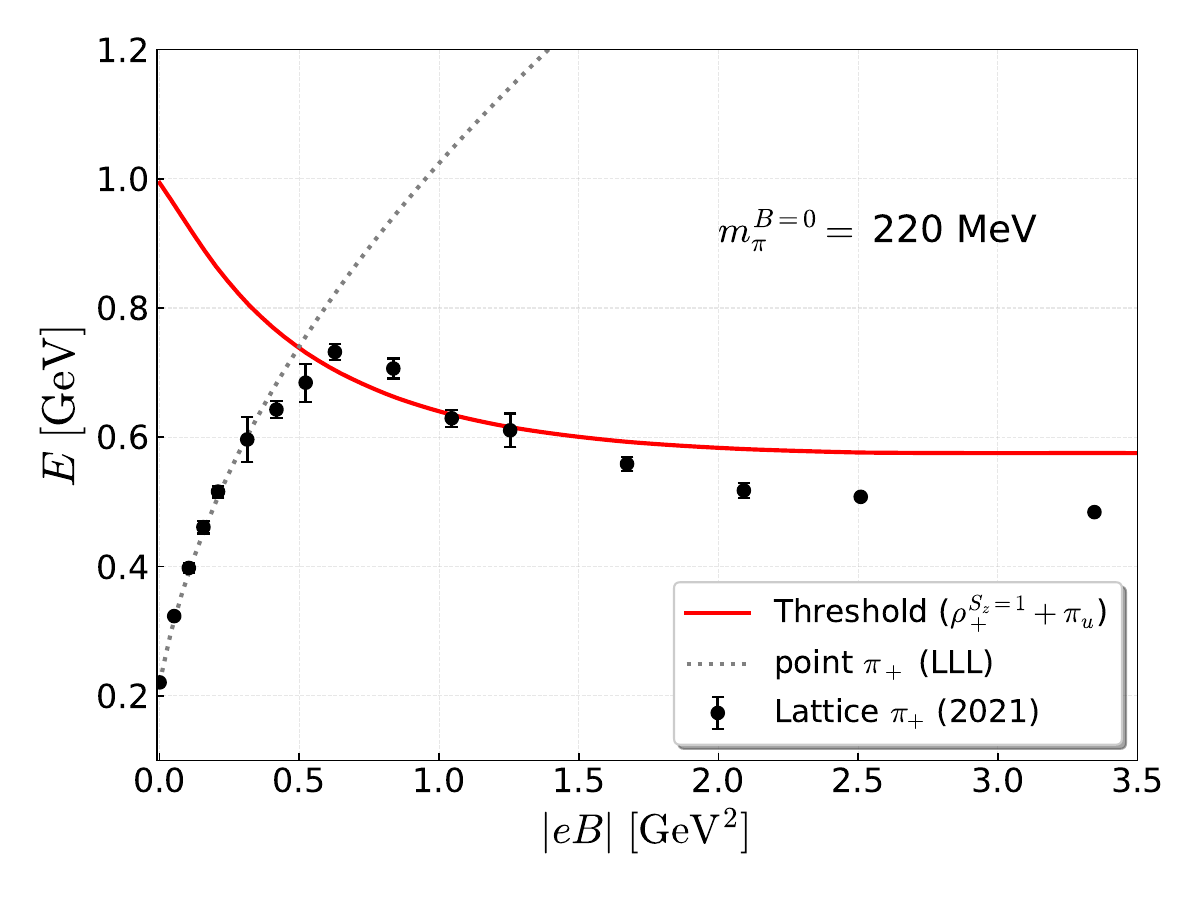}
\end{center}
\vspace{-0.8cm}
\caption{
The energy spectrum of $\pi_+$ in full QCD lattice simulations
with the pion mass $m_\pi^{B=0}\simeq 220$ MeV~\cite{Ding:2020hxw}.
The two-meson threshold $\pi_u+\rho_+^{S_z=1}$ (solid) and a point-particle estimate for $\pi_+$ (dotted) are also shown.
The energy of $\pi_u\sim u\bar u$ is taken from Ref.~\cite{Ding:2020hxw},
while that of $\rho_+^{S_z=1}$ is estimated from the $\sim 30\%$ energy reduction
observed in the lattice data of Ref.~\cite{Bali:2017ian}.
}
\vspace{-0.3cm}
\label{fig:threshold_pi220}
\end{figure}   

\section{Implications from lattice} \label{sec:lattice}

Lattice simulations in Fig.~\ref{fig:threshold_pi220} show that the $\pi_+$ energy
turns over and approaches a nearly constant value at large $B$.
The corresponding two-meson thresholds,
constructed from $\pi_u$ and vector mesons ($\rho_+^{S_z=1}$), lie close to the observed spectra.
The finite-$B$ energies of the constituent mesons used to construct the threshold are taken from lattice calculations;
they show the energy reduction by $\sim$ 30\% at $eB \simeq 1\,{\rm GeV}^2$ \cite{Bali:2017ian,Ding:2026qzu}.
Furthermore, the extracted energies exhibit only weak volume dependence \cite{Ding:2020hxw}, 
suggesting that the corresponding states are spatially localized.
Although not shown here, a similar tendency is also seen in the $K_+$ channel,
whose large-$B$ spectra also lie close to the two-meson threshold $K_{*+}^{S_z=1}+\pi_u$ \cite{Ding:2020hxw,Ding:2026qzu}.

These observations motivate us to investigate the low-lying states within a two-meson framework.

\section{Hamiltonian for two mesons} \label{sec:hamiltonian_2mesons}

We consider a two-meson system composed of a neutral meson and a charged meson.
The Hamiltonian is
\beq
H
& =
 H_{\rm n} + H_{\rm c} + U_{\rm nc} (\bm{R}_{\rm n} - \bm{R}_{\rm c} ) \,,
\eeq 
where $ H_{\rm n}$,  $H_{\rm c} $, and $U_{\rm nc} $ denote
Hamiltonians for a neutral meson, a charged meson, and the interaction potential between these two mesons, respectively.
The coordinates $\bm{R}_{\rm n}$ and $\bm{R}_{\rm c}$ are the center-of-mass positions of the neutral and charged mesons.
We first summarize the effective dynamics of $H_{\rm n}$ and $H_{\rm c}$ separately, 
and then combine them to derive the Born-Oppenheimer equation for the two-meson system.

To develop analytic insights, we consider a non-relativistic quark model \cite{DeRujula:1975qlm,Isgur:1979be}
\beq
H_M (e_1,e_2)
=
\sum_{i=1,2}
\frac{\, {\bm \Pi}_i^2-e_i \sigma_z B \,} {2m}
+V (\br) \,,
\eeq
where $\bm{\Pi}_i =  {\bm p}_i - e_i {\bm A}_i$ and $\bm{r} = \br_1 - \br_2$, and we have taken equal quark masses $m$.
The potential $V(\bm r)$ represents an effective quark-antiquark interaction,
including confining and one-gluon-exchange components,
whose detailed form is not essential for the dimensional-reduction mechanism discussed below.
For neutral mesons with $(e_1,e_2)=(-q,q)$ and charged mesons with
$(e_1,e_2)=(2/3,1/3)e$,
we define
\beq
H_{\rm n}\equiv H_M(-q,q) \,,
\qquad
H_{\rm c}\equiv H_M(2e/3,e/3)\,.
\eeq
The key difference between neutral and charged mesons
originates from the algebra of the pseudomomentum,
\beq
\bm{\calK}
=
\sum_{i=1,2}
\big(
\bm{\Pi}_i + e_i \bm{B} \times \bm{r}_i
\big) \,,
\eeq
whose transverse components satisfy
\beq
[\calK_x, \calK_y]
=
- \rmi Q B,
\qquad
Q = e_1+e_2 \,.
\eeq
Thus, $\calK_x$ and $\calK_y$ commute for neutral mesons ($Q=0$),
whereas they do not for charged mesons ($Q \neq0$).

With the Born-Oppenheimer construction in mind,
we first study neutral and charged mesons in a weak external potential
representing the interaction with the other meson.

\section{Neutral pions in an external potential} \label{sec:neutral}

We consider a neutral meson in an external potential $U_{\rm ex}(\bm R)$.
The corresponding problem for $U_{\rm ex}=0$ has been solved previously \cite{Alford:2013jva,Andreichikov:2016ayj,Yoshida:2016xgm,Cao:2025rop}.
Here we summarize only the ingredients needed for the Born-Oppenheimer description.

For $U_{\rm ex} = 0$, the pseudomomentum $\bm{\calK}$ is conserved
and its components commute, $[\calK_x,\calK_y]=0$.
The corresponding eigenstates are labeled by $\bm K$.
After a unitary transformation \cite{Kojo:2026gep}, the center-of-mass coordinate $\bm{R}_\perp$ is eliminated 
and the transverse pseudomomentum $\bm{K}_\perp$ enters only through a shift of the interaction potential,
\beq
V(\bm r)
\rightarrow
V(\bm r+\bm r_{\bm K}),
\eeq
where $\bm{r}_{\bm K} = \hat{z} \times \bm{K}_\perp /qB$.
The shift $\bm r_{\bm K}$ grows linearly with $\bm{K}_\perp$ but is suppressed by the magnetic field.
Consequently, the dependence of the meson energy on $\bm{K}_\perp$ becomes increasingly weak at large $B$.

Now we consider $U_{\rm ex} \neq 0$.
The eigenvalue $\epsilon_{\rm n} (\bm{K})$ for the fast motion
is obtained from
\beq
 H_{\rm n} \varphi_{\bm{K}} ( \br ) 
= \epsilon_{\rm n} (\bm{K} ) \varphi_{\bm{K}} ( \br ) 
\,.
\eeq
The corresponding equation for the slow center-of-mass motion is
\begin{align}
\big[ E - \epsilon_{\rm n} (\bm{K}) \big] \Phi (\bm{K})
=
 \int_{\bm{K'} } \tilde{U}_{\rm eff} (\bm{K}, \bm{K}')  \Phi (\bm{K}') \,,
 \label{eq:eigen_full}
\end{align}
where the effective potential  is
\beq
 \tilde{U}_{\rm eff} (\bm{K}, \bm{K}') 
 = \tilde{U}_{\rm ex} (\bm{K} - \bm{K}') 
 \int_{\br} 
\varphi^*_{\bm{K}} ( \br ) 
\varphi_{\bm{K}'} ( \br ) 
 \,.
\eeq
For $K_\perp,K'_\perp\sim\lqcd$ at large $B$, the shifts
$\bm{r}_{\bm K}$ and $\bm{r}_{\bm K'}$ are small, so that
$\tilde U_{\rm eff}(\bm K,\bm K')\simeq \tilde U_{\rm ex}(\bm K-\bm K')$.
Equation~\eqref{eq:eigen_full} can then be Fourier transformed to coordinate space,
\beq
\big[
E-\epsilon_{\rm n}(-\rmi \bm{\nabla}_{R} ) - U_{\rm ex}({\bm R})
\big] 
\Phi({\bm R})
=0 \, .
\eeq
The dimensional reduction follows from the expansion of $\epsilon_{\rm n}$ in powers of $\bm K$,
\beq
 \epsilon_{\rm n} (\bm{K} )
 \simeq  
 \epsilon_{\rm n} (0)
+ \frac{\, K_z^2 \,}{\, 2M\, } + \frac{\, K_\perp^2 \,}{\, 2M_B\, } + \cdots
\eeq
where $M=2m$, and $M_B$ is a transverse effective mass that grows with $B$.
For a Coulomb-like short-distance potential one finds
$M_B \sim B/\lqcd$~\cite{Kojo:2026gep},
while the precise scaling is model dependent \cite{Hattori:2015aki}.
The essential point is that $M_B$ becomes parametrically large at strong magnetic fields.

The suppression of the transverse kinetic term by $1/M_B$ implies that
$\bm R_\perp$ can be treated semiclassically and fixed near the minimum of
$U_{\rm ex}$ for a given $R_z$,
since quantum fluctuations around this minimum are suppressed.
As a result, the slow collective dynamics is effectively one-dimensional and governed by the coordinate $R_z$.

\section{ Charged mesons in an external potential} \label{sec:qqbar_charged}

Next we consider a charged meson in an external potential $U_{\rm ex}(\bm R)$.
For $U_{\rm ex}=0$, the pseudomomentum components do not commute,
$[\calK_x,\calK_y] = -\rmi QB$.
Choosing $K_x$ to label the states,
one finds that the transverse dynamics is quantized into Landau levels and the spectrum is independent of $K_x$ \cite{Kojo:2026gep}.

Now we consider $U_{\rm ex} \neq 0$.
Below we use the notation $\tilde{\bm{K}}\equiv (K_x, K_z)$.
After a unitary transformation \cite{Kojo:2026gep},
the center-of-mass coordinate enters through the shifted combination
$\bm R-\hat y K_x/QB$.
We denote the corresponding shifted Hamiltonian by $H_{\rm c}^{K_x}$.
Keeping the longitudinal plane wave explicit, the eigenvalue $\epsilon_{\rm c}(\tilde{\bm K})$ for the fast motion is obtained from
\beq
 \big[\, H_{\rm c}^{ K_x } - \epsilon_{\rm c}(\tilde{\bm K}) \, \big] \rme^{\rmi K_z R_z} \varphi_{ K_x } (\br, R_y) 
 = 0 \,.
\eeq
The corresponding equation for the slow center-of-mass motion is
\begin{align}
\big[ E - \epsilon_{\rm c}(\tilde{\bm K}) \big] \Phi (\tilde{\bm K})
=
 \int_{ \tilde{\bm K}' } \tilde{U}_{\rm eff} (\tilde{\bm K},\tilde{\bm K}' )  \Phi ( \tilde{\bm K}') \,,
 \label{eq:eigen_full_ch}
\end{align}
where the effective potential  is
\begin{align}
& \tilde{U}_{\rm eff} (\tilde{\bm K},\tilde{\bm K}') 
\notag \\
&\!\! = \int_{\br, \bm{R} } 
\rme^{ \rmi ( \tilde{\bm K} - \tilde{\bm K}' ) \cdot \tilde{\bm R} } 
\, \varphi_{K_x}^* (\br, R_y ) 
U_{\rm ex} (\bm{R} ) 
 \varphi_{K'_x} (\br, R_y ) 
 \,.
\end{align}
The coordinate $R_y$ is already incorporated into the fast wave function
$\varphi_{K_x} (\bm r,R_y)$ and remains localized around $R_y \simeq 0$ within a Landau orbit of the radius $\sim B^{-1/2}$.

For $K_x,K_x'\sim\lqcd$ at large $B$,  the residual $K_x$ dependence of
$\epsilon_{\rm c}(\tilde{\bm K})$ and of the overlap factor may be neglected.
Fourier transforming Eq.~\eqref{eq:eigen_full_ch}, we obtain
\beq
\big[
E-\epsilon_{\rm c}(-\rmi\partial_{R_z}) - U_{\rm eff}(\tilde{\bm R})
\big] 
\Phi(\tilde{\bm R})
=0 \, .
\eeq
The $R_x$ dependence is generated only by the external potential, and there is no transverse kinetic term for $R_x$.
As in the neutral meson case, $R_x$ can therefore be treated semiclassically.
The slow collective dynamics is then effectively one-dimensional and governed by the coordinate $R_z$.

\section{ Meson molecules } \label{sec:meson_molecules}

As shown in the previous sections,
the transverse quantum fluctuations of both neutral and charged mesons are strongly suppressed at large magnetic fields.
Defining $\bm{R}_{\rm nc} \equiv \bm{R}_{\rm n} - \bm{R}_{\rm c}$,
the meson-meson interaction can be evaluated at the expectation values of the collective coordinates,
\beq
U_{\rm nc}(\bm R_{\rm nc})
\rightarrow
U_{\rm nc}
\big(
\la \bm R_{\rm nc}^\perp \ra, R_{\rm nc}^z  
\big) 
\equiv U_{\rm BO} (R_{\rm nc}^z)
\,,
\eeq
where $\la \bm{R}_{\rm nc}^{\perp} \ra$ is determined by minimizing the effective interaction at fixed $R_{\rm nc}^{z}$.
The resulting Born-Oppenheimer potential $U_{\rm BO}(R_{\rm nc}^{z})$ determines the molecular spectrum.

The validity of this replacement depends on the structure of the interaction.
For regular hadronic interactions characterized by the length scales
$m_\pi^{-1}$ or $\Lambda_{\rm QCD}^{-1}$,
the suppression of transverse dynamics discussed above
results in transverse quantum fluctuations $\Delta R_\perp$
much smaller than these scales.
In this case the dimensional reduction becomes increasingly accurate at large $B$.
For a singular short-distance attraction such as $U_{\rm nc}(R)\sim 1/R$,
however, the gain in potential energy at short distances can overcome the magnetic suppression of transverse dynamics.
The system is then driven toward arbitrarily small separations
and the transverse dynamics must be solved explicitly.
In this case the system enters the compact multiquark regime.

The emergence of molecular bound states can be understood from
the Lippmann-Schwinger equation,
\beq
\hspace{-0.5cm}
\Phi (K_z)
=
\frac{1}{\, E_{\rm b} + K_z^2/2\mu_{\rm nc} \,}
\int_{K_z'} \!
\tilde U_{\rm BO}(K_z-K_z')\,
\Phi(K_z') ,
\eeq
where $\mu_{\rm nc}^{-1} = m_{\rm n}^{-1} + m_{\rm c}^{-1}$ is the reduced mass, 
and $E_{\rm b}$ is the binding energy.
In the low energy limit, we expand the potential in powers of momenta,
$\tilde U_{\rm BO}(K_z-K_z') \simeq -G_0 + O[(K_z-K_z')^2]$.
Defining $C_0 \equiv \int_{K_z}\Phi(K_z)$ and integrating both sides over $K_z$, 
we obtain the equation to determine the bound state pole
\beq
1 \simeq \int_{K_z} \frac{\, G_0 \,}{\, E_{\rm b} +K_z^2/2\mu_{\rm nc} \,} 
~\rightarrow~
E_{\rm b} \simeq \frac{\, \mu_{\rm nc} G_0^2 \,}{\, 2 \,}
\,.
\eeq
The special feature of one spatial dimension is that the RHS diverges in the infrared at $E_{\rm b}=0$.
As $E_{\rm b}$ increases, the RHS decreases continuously from infinity to zero.
Therefore, a solution always exists for $G_0>0$.
This condition is equivalent to $\int \rmd R_z\,U_{\rm BO}(R_z) <0$,
i.e., the net interaction is attractive.

\section{Summary} \label{sec:summary}

We have shown that the effective dynamics of neutral and charged mesons in strong magnetic fields naturally leads to a Born-Oppenheimer description of two-meson systems.
For both neutral and charged mesons,
the transverse quantum fluctuations are strongly suppressed,
and hence the meson motion becomes effectively one-dimensional,
allowing a dimensional reduction of the meson-meson problem.
A characteristic feature of one dimension is that
any net attractive interaction generates a bound state.
Therefore strong magnetic fields promote the formation
of mesonic molecules even when the attraction is weak.

This mechanism provides a natural explanation for the
non-monotonic behavior observed in lattice simulations of
charged pions and kaons \cite{Ding:2020hxw,Ding:2026qzu}.
The energies extracted on the lattice lie close to the
two-meson thresholds, supporting the relevance of a molecular description.
The framework developed here also clarifies the boundary
between molecular and compact multiquark regimes in strong magnetic fields.

In this paper we consider the combination of neutral and charged mesons
but other combinations, including neutral-neutral and charged-charged mesons, 
as well as meson-baryon molecular states, are also possible. 
Further studies along these lines may provide
a systematic framework for understanding the interplay
between molecular and compact multiquark structures
in strong magnetic fields,
with possible implications for exotic hadrons at $B=0$.
\\
\\

\begin{acknowledgments}
I thank Gaoqing Cao, Sakura Itatani, and Heng-Tong Ding for discussions at the initial stage of this work,
and Dan Zhang, Gergely Endr{\H{o}}di, and Bastian Brandt for sharing their data with me.
This work is supported by
JSPS KAKENHI Grant No. 26K07077.
%
\end{acknowledgments}

\bibliography{ref}

\begin{thebibliography}{24}%
\makeatletter
\providecommand \@ifxundefined [1]{%
 \@ifx{#1\undefined}
}%
\providecommand \@ifnum [1]{%
 \ifnum #1\expandafter \@firstoftwo
 \else \expandafter \@secondoftwo
 \fi
}%
\providecommand \@ifx [1]{%
 \ifx #1\expandafter \@firstoftwo
 \else \expandafter \@secondoftwo
 \fi
}%
\providecommand \natexlab [1]{#1}%
\providecommand \enquote  [1]{``#1''}%
\providecommand \bibnamefont  [1]{#1}%
\providecommand \bibfnamefont [1]{#1}%
\providecommand \citenamefont [1]{#1}%
\providecommand \href@noop [0]{\@secondoftwo}%
\providecommand \href [0]{\begingroup \@sanitize@url \@href}%
\providecommand \@href[1]{\@@startlink{#1}\@@href}%
\providecommand \@@href[1]{\endgroup#1\@@endlink}%
\providecommand \@sanitize@url [0]{\catcode `\\12\catcode `\$12\catcode
  `\&12\catcode `\#12\catcode `\^12\catcode `\_12\catcode `\%12\relax}%
\providecommand \@@startlink[1]{}%
\providecommand \@@endlink[0]{}%
\providecommand \url  [0]{\begingroup\@sanitize@url \@url }%
\providecommand \@url [1]{\endgroup\@href {#1}{\urlprefix }}%
\providecommand \urlprefix  [0]{URL }%
\providecommand \Eprint [0]{\href }%
\providecommand \doibase [0]{https://doi.org/}%
\providecommand \selectlanguage [0]{\@gobble}%
\providecommand \bibinfo  [0]{\@secondoftwo}%
\providecommand \bibfield  [0]{\@secondoftwo}%
\providecommand \translation [1]{[#1]}%
\providecommand \BibitemOpen [0]{}%
\providecommand \bibitemStop [0]{}%
\providecommand \bibitemNoStop [0]{.\EOS\space}%
\providecommand \EOS [0]{\spacefactor3000\relax}%
\providecommand \BibitemShut  [1]{\csname bibitem#1\endcsname}%
\let\auto@bib@innerbib\@empty
\bibitem [{\citenamefont {Guo}\ \emph {et~al.}(2018)\citenamefont {Guo},
  \citenamefont {Hanhart}, \citenamefont {Mei{\ss}ner}, \citenamefont {Wang},
  \citenamefont {Zhao},\ and\ \citenamefont {Zou}}]{Guo:2017jvc}%
  \BibitemOpen
  \bibfield  {author} {\bibinfo {author} {\bibfnamefont {F.-K.}\ \bibnamefont
  {Guo}}, \bibinfo {author} {\bibfnamefont {C.}~\bibnamefont {Hanhart}},
  \bibinfo {author} {\bibfnamefont {U.-G.}\ \bibnamefont {Mei{\ss}ner}},
  \bibinfo {author} {\bibfnamefont {Q.}~\bibnamefont {Wang}}, \bibinfo {author}
  {\bibfnamefont {Q.}~\bibnamefont {Zhao}},\ and\ \bibinfo {author}
  {\bibfnamefont {B.-S.}\ \bibnamefont {Zou}},\ }\bibfield  {title} {\bibinfo
  {title} {{Hadronic molecules}},\ }\href
  {https://doi.org/10.1103/RevModPhys.90.015004} {\bibfield  {journal}
  {\bibinfo  {journal} {Rev. Mod. Phys.}\ }\textbf {\bibinfo {volume} {90}},\
  \bibinfo {pages} {015004} (\bibinfo {year} {2018})},\ \bibinfo {note}
  {[Erratum: Rev.Mod.Phys. 94, 029901 (2022)]},\ \Eprint
  {https://arxiv.org/abs/1705.00141} {arXiv:1705.00141 [hep-ph]} \BibitemShut
  {NoStop}%
\bibitem [{\citenamefont {Esposito}\ \emph {et~al.}(2017)\citenamefont
  {Esposito}, \citenamefont {Pilloni},\ and\ \citenamefont
  {Polosa}}]{Esposito:2016noz}%
  \BibitemOpen
  \bibfield  {author} {\bibinfo {author} {\bibfnamefont {A.}~\bibnamefont
  {Esposito}}, \bibinfo {author} {\bibfnamefont {A.}~\bibnamefont {Pilloni}},\
  and\ \bibinfo {author} {\bibfnamefont {A.~D.}\ \bibnamefont {Polosa}},\
  }\bibfield  {title} {\bibinfo {title} {{Multiquark Resonances}},\ }\href
  {https://doi.org/10.1016/j.physrep.2016.11.002} {\bibfield  {journal}
  {\bibinfo  {journal} {Phys. Rept.}\ }\textbf {\bibinfo {volume} {668}},\
  \bibinfo {pages} {1} (\bibinfo {year} {2017})},\ \Eprint
  {https://arxiv.org/abs/1611.07920} {arXiv:1611.07920 [hep-ph]} \BibitemShut
  {NoStop}%
\bibitem [{\citenamefont {Lebed}\ \emph {et~al.}(2017)\citenamefont {Lebed},
  \citenamefont {Mitchell},\ and\ \citenamefont {Swanson}}]{Lebed:2016hpi}%
  \BibitemOpen
  \bibfield  {author} {\bibinfo {author} {\bibfnamefont {R.~F.}\ \bibnamefont
  {Lebed}}, \bibinfo {author} {\bibfnamefont {R.~E.}\ \bibnamefont
  {Mitchell}},\ and\ \bibinfo {author} {\bibfnamefont {E.~S.}\ \bibnamefont
  {Swanson}},\ }\bibfield  {title} {\bibinfo {title} {{Heavy-Quark QCD
  Exotica}},\ }\href {https://doi.org/10.1016/j.ppnp.2016.11.003} {\bibfield
  {journal} {\bibinfo  {journal} {Prog. Part. Nucl. Phys.}\ }\textbf {\bibinfo
  {volume} {93}},\ \bibinfo {pages} {143} (\bibinfo {year} {2017})},\ \Eprint
  {https://arxiv.org/abs/1610.04528} {arXiv:1610.04528 [hep-ph]} \BibitemShut
  {NoStop}%
\bibitem [{\citenamefont {Huang}(2016)}]{Huang:2015oca}%
  \BibitemOpen
  \bibfield  {author} {\bibinfo {author} {\bibfnamefont {X.-G.}\ \bibnamefont
  {Huang}},\ }\bibfield  {title} {\bibinfo {title} {{Electromagnetic fields and
  anomalous transports in heavy-ion collisions --- A pedagogical review}},\
  }\href@noop {} {\bibfield  {journal} {\bibinfo  {journal} {Rept. Prog.
  Phys.}\ }\textbf {\bibinfo {volume} {79}},\ \bibinfo {pages} {076302}
  (\bibinfo {year} {2016})},\ \Eprint {https://arxiv.org/abs/1509.04073}
  {arXiv:1509.04073 [nucl-th]} \BibitemShut {NoStop}%
\bibitem [{\citenamefont {Bali}\ \emph {et~al.}(2018)\citenamefont {Bali},
  \citenamefont {Brandt}, \citenamefont {Endr{\H{o}}di},\ and\ \citenamefont
  {Gl{\"a}{\ss}le}}]{Bali:2017ian}%
  \BibitemOpen
  \bibfield  {author} {\bibinfo {author} {\bibfnamefont {G.~S.}\ \bibnamefont
  {Bali}}, \bibinfo {author} {\bibfnamefont {B.~B.}\ \bibnamefont {Brandt}},
  \bibinfo {author} {\bibfnamefont {G.}~\bibnamefont {Endr{\H{o}}di}},\ and\
  \bibinfo {author} {\bibfnamefont {B.}~\bibnamefont {Gl{\"a}{\ss}le}},\
  }\bibfield  {title} {\bibinfo {title} {{Meson masses in electromagnetic
  fields with Wilson fermions}},\ }\href
  {https://doi.org/10.1103/PhysRevD.97.034505} {\bibfield  {journal} {\bibinfo
  {journal} {Phys. Rev. D}\ }\textbf {\bibinfo {volume} {97}},\ \bibinfo
  {pages} {034505} (\bibinfo {year} {2018})},\ \Eprint
  {https://arxiv.org/abs/1707.05600} {arXiv:1707.05600 [hep-lat]} \BibitemShut
  {NoStop}%
\bibitem [{\citenamefont {D'Elia}\ \emph {et~al.}(2021)\citenamefont {D'Elia},
  \citenamefont {Maio}, \citenamefont {Sanfilippo},\ and\ \citenamefont
  {Stanzione}}]{DElia:2021tfb}%
  \BibitemOpen
  \bibfield  {author} {\bibinfo {author} {\bibfnamefont {M.}~\bibnamefont
  {D'Elia}}, \bibinfo {author} {\bibfnamefont {L.}~\bibnamefont {Maio}},
  \bibinfo {author} {\bibfnamefont {F.}~\bibnamefont {Sanfilippo}},\ and\
  \bibinfo {author} {\bibfnamefont {A.}~\bibnamefont {Stanzione}},\ }\bibfield
  {title} {\bibinfo {title} {{Confining and chiral properties of QCD in
  extremely strong magnetic fields}},\ }\href
  {https://doi.org/10.1103/PhysRevD.104.114512} {\bibfield  {journal} {\bibinfo
   {journal} {Phys. Rev. D}\ }\textbf {\bibinfo {volume} {104}},\ \bibinfo
  {pages} {114512} (\bibinfo {year} {2021})},\ \Eprint
  {https://arxiv.org/abs/2109.07456} {arXiv:2109.07456 [hep-lat]} \BibitemShut
  {NoStop}%
\bibitem [{\citenamefont {Andersen}\ \emph {et~al.}(2016)\citenamefont
  {Andersen}, \citenamefont {Naylor},\ and\ \citenamefont
  {Tranberg}}]{Andersen:2014xxa}%
  \BibitemOpen
  \bibfield  {author} {\bibinfo {author} {\bibfnamefont {J.~O.}\ \bibnamefont
  {Andersen}}, \bibinfo {author} {\bibfnamefont {W.~R.}\ \bibnamefont
  {Naylor}},\ and\ \bibinfo {author} {\bibfnamefont {A.}~\bibnamefont
  {Tranberg}},\ }\bibfield  {title} {\bibinfo {title} {{Phase diagram of QCD in
  a magnetic field: A review}},\ }\href
  {https://doi.org/10.1103/RevModPhys.88.025001} {\bibfield  {journal}
  {\bibinfo  {journal} {Rev. Mod. Phys.}\ }\textbf {\bibinfo {volume} {88}},\
  \bibinfo {pages} {025001} (\bibinfo {year} {2016})},\ \Eprint
  {https://arxiv.org/abs/1411.7176} {arXiv:1411.7176 [hep-ph]} \BibitemShut
  {NoStop}%
\bibitem [{\citenamefont {Miransky}\ and\ \citenamefont
  {Shovkovy}(2015)}]{Miransky:2015ava}%
  \BibitemOpen
  \bibfield  {author} {\bibinfo {author} {\bibfnamefont {V.~A.}\ \bibnamefont
  {Miransky}}\ and\ \bibinfo {author} {\bibfnamefont {I.~A.}\ \bibnamefont
  {Shovkovy}},\ }\bibfield  {title} {\bibinfo {title} {{Quantum field theory in
  a magnetic field: From quantum chromodynamics to graphene and Dirac
  semimetals}},\ }\href {https://doi.org/10.1016/j.physrep.2015.02.003}
  {\bibfield  {journal} {\bibinfo  {journal} {Phys. Rept.}\ }\textbf {\bibinfo
  {volume} {576}},\ \bibinfo {pages} {1} (\bibinfo {year} {2015})},\ \Eprint
  {https://arxiv.org/abs/1503.00732} {arXiv:1503.00732 [hep-ph]} \BibitemShut
  {NoStop}%
\bibitem [{\citenamefont {Fukushima}\ and\ \citenamefont
  {Hidaka}(2013)}]{Fukushima:2012kc}%
  \BibitemOpen
  \bibfield  {author} {\bibinfo {author} {\bibfnamefont {K.}~\bibnamefont
  {Fukushima}}\ and\ \bibinfo {author} {\bibfnamefont {Y.}~\bibnamefont
  {Hidaka}},\ }\bibfield  {title} {\bibinfo {title} {{Magnetic Catalysis Versus
  Magnetic Inhibition}},\ }\href
  {https://doi.org/10.1103/PhysRevLett.110.031601} {\bibfield  {journal}
  {\bibinfo  {journal} {Phys. Rev. Lett.}\ }\textbf {\bibinfo {volume} {110}},\
  \bibinfo {pages} {031601} (\bibinfo {year} {2013})},\ \Eprint
  {https://arxiv.org/abs/1209.1319} {arXiv:1209.1319 [hep-ph]} \BibitemShut
  {NoStop}%
\bibitem [{\citenamefont {Kojo}(2021)}]{Kojo:2021gvm}%
  \BibitemOpen
  \bibfield  {author} {\bibinfo {author} {\bibfnamefont {T.}~\bibnamefont
  {Kojo}},\ }\bibfield  {title} {\bibinfo {title} {{Neutral and charged mesons
  in magnetic fields: A resonance gas in a non-relativistic quark model}},\
  }\href {https://doi.org/10.1140/epja/s10050-021-00629-y} {\bibfield
  {journal} {\bibinfo  {journal} {Eur. Phys. J. A}\ }\textbf {\bibinfo {volume}
  {57}},\ \bibinfo {pages} {317} (\bibinfo {year} {2021})},\ \Eprint
  {https://arxiv.org/abs/2104.00376} {arXiv:2104.00376 [hep-ph]} \BibitemShut
  {NoStop}%
\bibitem [{\citenamefont {Ding}\ \emph {et~al.}(2021)\citenamefont {Ding},
  \citenamefont {Li}, \citenamefont {Tomiya}, \citenamefont {Wang},\ and\
  \citenamefont {Zhang}}]{Ding:2020hxw}%
  \BibitemOpen
  \bibfield  {author} {\bibinfo {author} {\bibfnamefont {H.~T.}\ \bibnamefont
  {Ding}}, \bibinfo {author} {\bibfnamefont {S.~T.}\ \bibnamefont {Li}},
  \bibinfo {author} {\bibfnamefont {A.}~\bibnamefont {Tomiya}}, \bibinfo
  {author} {\bibfnamefont {X.~D.}\ \bibnamefont {Wang}},\ and\ \bibinfo
  {author} {\bibfnamefont {Y.}~\bibnamefont {Zhang}},\ }\bibfield  {title}
  {\bibinfo {title} {{Chiral properties of (2+1)-flavor QCD in strong magnetic
  fields at zero temperature}},\ }\href
  {https://doi.org/10.1103/PhysRevD.104.014505} {\bibfield  {journal} {\bibinfo
   {journal} {Phys. Rev. D}\ }\textbf {\bibinfo {volume} {104}},\ \bibinfo
  {pages} {014505} (\bibinfo {year} {2021})},\ \Eprint
  {https://arxiv.org/abs/2008.00493} {arXiv:2008.00493 [hep-lat]} \BibitemShut
  {NoStop}%
\bibitem [{\citenamefont {Ding}\ and\ \citenamefont
  {Zhang}(2026)}]{Ding:2026qzu}%
  \BibitemOpen
  \bibfield  {author} {\bibinfo {author} {\bibfnamefont {H.-T.}\ \bibnamefont
  {Ding}}\ and\ \bibinfo {author} {\bibfnamefont {D.}~\bibnamefont {Zhang}},\
  }\bibfield  {title} {\bibinfo {title} {{Chiral Properties of
  $(2\!+\!1)$-Flavor QCD in Magnetic Fields at Zero Temperature}},\ }\href@noop
  {} {\  (\bibinfo {year} {2026})},\ \Eprint {https://arxiv.org/abs/2601.18354}
  {arXiv:2601.18354 [hep-lat]} \BibitemShut {NoStop}%
\bibitem [{\citenamefont {Taya}(2015)}]{Taya:2014nha}%
  \BibitemOpen
  \bibfield  {author} {\bibinfo {author} {\bibfnamefont {H.}~\bibnamefont
  {Taya}},\ }\bibfield  {title} {\bibinfo {title} {{Hadron Masses in Strong
  Magnetic Fields}},\ }\href {https://doi.org/10.1103/PhysRevD.92.014038}
  {\bibfield  {journal} {\bibinfo  {journal} {Phys. Rev. D}\ }\textbf {\bibinfo
  {volume} {92}},\ \bibinfo {pages} {014038} (\bibinfo {year} {2015})},\
  \Eprint {https://arxiv.org/abs/1412.6877} {arXiv:1412.6877 [hep-ph]}
  \BibitemShut {NoStop}%
\bibitem [{\citenamefont {Kojo}\ and\ \citenamefont {Su}(2013)}]{Kojo:2012js}%
  \BibitemOpen
  \bibfield  {author} {\bibinfo {author} {\bibfnamefont {T.}~\bibnamefont
  {Kojo}}\ and\ \bibinfo {author} {\bibfnamefont {N.}~\bibnamefont {Su}},\
  }\bibfield  {title} {\bibinfo {title} {{The quark mass gap in a magnetic
  field}},\ }\href {https://doi.org/10.1016/j.physletb.2013.02.024} {\bibfield
  {journal} {\bibinfo  {journal} {Phys. Lett. B}\ }\textbf {\bibinfo {volume}
  {720}},\ \bibinfo {pages} {192} (\bibinfo {year} {2013})},\ \Eprint
  {https://arxiv.org/abs/1211.7318} {arXiv:1211.7318 [hep-ph]} \BibitemShut
  {NoStop}%
\bibitem [{\citenamefont {Bali}\ \emph {et~al.}(2012)\citenamefont {Bali},
  \citenamefont {Bruckmann}, \citenamefont {Endrodi}, \citenamefont {Fodor},
  \citenamefont {Katz}, \citenamefont {Krieg}, \citenamefont {Schafer},\ and\
  \citenamefont {Szabo}}]{Bali:2011qj}%
  \BibitemOpen
  \bibfield  {author} {\bibinfo {author} {\bibfnamefont {G.~S.}\ \bibnamefont
  {Bali}}, \bibinfo {author} {\bibfnamefont {F.}~\bibnamefont {Bruckmann}},
  \bibinfo {author} {\bibfnamefont {G.}~\bibnamefont {Endrodi}}, \bibinfo
  {author} {\bibfnamefont {Z.}~\bibnamefont {Fodor}}, \bibinfo {author}
  {\bibfnamefont {S.~D.}\ \bibnamefont {Katz}}, \bibinfo {author}
  {\bibfnamefont {S.}~\bibnamefont {Krieg}}, \bibinfo {author} {\bibfnamefont
  {A.}~\bibnamefont {Schafer}},\ and\ \bibinfo {author} {\bibfnamefont {K.~K.}\
  \bibnamefont {Szabo}},\ }\bibfield  {title} {\bibinfo {title} {{The QCD phase
  diagram for external magnetic fields}},\ }\href
  {https://doi.org/10.1007/JHEP02(2012)044} {\bibfield  {journal} {\bibinfo
  {journal} {JHEP}\ }\textbf {\bibinfo {volume} {02}},\ \bibinfo {pages}
  {044}},\ \Eprint {https://arxiv.org/abs/1111.4956} {arXiv:1111.4956
  [hep-lat]} \BibitemShut {NoStop}%
\bibitem [{\citenamefont {Wang}(2026)}]{Wang:2026xsm}%
  \BibitemOpen
  \bibfield  {author} {\bibinfo {author} {\bibfnamefont {Z.}~\bibnamefont
  {Wang}},\ }\bibfield  {title} {\bibinfo {title} {{Residue-Enhanced Pion-Rho
  Mixing as the Origin of Nonmonotonic Charged Pion Mass in Magnetic Fields}},\
  }\href@noop {} {\  (\bibinfo {year} {2026})},\ \Eprint
  {https://arxiv.org/abs/2602.15410} {arXiv:2602.15410 [hep-ph]} \BibitemShut
  {NoStop}%
\bibitem [{\citenamefont {De~Rujula}\ \emph {et~al.}(1975)\citenamefont
  {De~Rujula}, \citenamefont {Georgi},\ and\ \citenamefont
  {Glashow}}]{DeRujula:1975qlm}%
  \BibitemOpen
  \bibfield  {author} {\bibinfo {author} {\bibfnamefont {A.}~\bibnamefont
  {De~Rujula}}, \bibinfo {author} {\bibfnamefont {H.}~\bibnamefont {Georgi}},\
  and\ \bibinfo {author} {\bibfnamefont {S.~L.}\ \bibnamefont {Glashow}},\
  }\bibfield  {title} {\bibinfo {title} {{Hadron Masses in a Gauge Theory}},\
  }\href {https://doi.org/10.1103/PhysRevD.12.147} {\bibfield  {journal}
  {\bibinfo  {journal} {Phys. Rev. D}\ }\textbf {\bibinfo {volume} {12}},\
  \bibinfo {pages} {147} (\bibinfo {year} {1975})}\BibitemShut {NoStop}%
\bibitem [{\citenamefont {Isgur}\ and\ \citenamefont
  {Karl}(1979)}]{Isgur:1979be}%
  \BibitemOpen
  \bibfield  {author} {\bibinfo {author} {\bibfnamefont {N.}~\bibnamefont
  {Isgur}}\ and\ \bibinfo {author} {\bibfnamefont {G.}~\bibnamefont {Karl}},\
  }\bibfield  {title} {\bibinfo {title} {{Ground State Baryons in a Quark Model
  with Hyperfine Interactions}},\ }\href
  {https://doi.org/10.1103/PhysRevD.20.1191} {\bibfield  {journal} {\bibinfo
  {journal} {Phys. Rev. D}\ }\textbf {\bibinfo {volume} {20}},\ \bibinfo
  {pages} {1191} (\bibinfo {year} {1979})}\BibitemShut {NoStop}%
\bibitem [{\citenamefont {Alford}\ and\ \citenamefont
  {Strickland}(2013)}]{Alford:2013jva}%
  \BibitemOpen
  \bibfield  {author} {\bibinfo {author} {\bibfnamefont {J.}~\bibnamefont
  {Alford}}\ and\ \bibinfo {author} {\bibfnamefont {M.}~\bibnamefont
  {Strickland}},\ }\bibfield  {title} {\bibinfo {title} {{Charmonia and
  Bottomonia in a Magnetic Field}},\ }\href
  {https://doi.org/10.1103/PhysRevD.88.105017} {\bibfield  {journal} {\bibinfo
  {journal} {Phys. Rev. D}\ }\textbf {\bibinfo {volume} {88}},\ \bibinfo
  {pages} {105017} (\bibinfo {year} {2013})},\ \Eprint
  {https://arxiv.org/abs/1309.3003} {arXiv:1309.3003 [hep-ph]} \BibitemShut
  {NoStop}%
\bibitem [{\citenamefont {Andreichikov}\ \emph {et~al.}(2017)\citenamefont
  {Andreichikov}, \citenamefont {Kerbikov}, \citenamefont {Luschevskaya},
  \citenamefont {Simonov},\ and\ \citenamefont
  {Solovjeva}}]{Andreichikov:2016ayj}%
  \BibitemOpen
  \bibfield  {author} {\bibinfo {author} {\bibfnamefont {M.~A.}\ \bibnamefont
  {Andreichikov}}, \bibinfo {author} {\bibfnamefont {B.~O.}\ \bibnamefont
  {Kerbikov}}, \bibinfo {author} {\bibfnamefont {E.~V.}\ \bibnamefont
  {Luschevskaya}}, \bibinfo {author} {\bibfnamefont {Y.~A.}\ \bibnamefont
  {Simonov}},\ and\ \bibinfo {author} {\bibfnamefont {O.~E.}\ \bibnamefont
  {Solovjeva}},\ }\bibfield  {title} {\bibinfo {title} {{The Evolution of Meson
  Masses in a Strong Magnetic Field}},\ }\href
  {https://doi.org/10.1007/JHEP05(2017)007} {\bibfield  {journal} {\bibinfo
  {journal} {JHEP}\ }\textbf {\bibinfo {volume} {05}},\ \bibinfo {pages}
  {007}},\ \Eprint {https://arxiv.org/abs/1610.06887} {arXiv:1610.06887
  [hep-ph]} \BibitemShut {NoStop}%
\bibitem [{\citenamefont {Yoshida}\ and\ \citenamefont
  {Suzuki}(2016)}]{Yoshida:2016xgm}%
  \BibitemOpen
  \bibfield  {author} {\bibinfo {author} {\bibfnamefont {T.}~\bibnamefont
  {Yoshida}}\ and\ \bibinfo {author} {\bibfnamefont {K.}~\bibnamefont
  {Suzuki}},\ }\bibfield  {title} {\bibinfo {title} {{Heavy meson spectroscopy
  under strong magnetic field}},\ }\href
  {https://doi.org/10.1103/PhysRevD.94.074043} {\bibfield  {journal} {\bibinfo
  {journal} {Phys. Rev. D}\ }\textbf {\bibinfo {volume} {94}},\ \bibinfo
  {pages} {074043} (\bibinfo {year} {2016})},\ \Eprint
  {https://arxiv.org/abs/1607.04935} {arXiv:1607.04935 [hep-ph]} \BibitemShut
  {NoStop}%
\bibitem [{\citenamefont {Cao}\ and\ \citenamefont {Wu}(2025)}]{Cao:2025rop}%
  \BibitemOpen
  \bibfield  {author} {\bibinfo {author} {\bibfnamefont {G.}~\bibnamefont
  {Cao}}\ and\ \bibinfo {author} {\bibfnamefont {S.}~\bibnamefont {Wu}},\
  }\bibfield  {title} {\bibinfo {title} {{Single-flavor heavy baryons in a
  strong magnetic field}},\ }\href {https://doi.org/10.1103/tr2g-7zxq}
  {\bibfield  {journal} {\bibinfo  {journal} {Phys. Rev. D}\ }\textbf {\bibinfo
  {volume} {112}},\ \bibinfo {pages} {074034} (\bibinfo {year} {2025})},\
  \Eprint {https://arxiv.org/abs/2507.20636} {arXiv:2507.20636 [nucl-th]}
  \BibitemShut {NoStop}%
\bibitem [{\citenamefont {Kojo}\ and\ \citenamefont
  {Itatani}(2026)}]{Kojo:2026gep}%
  \BibitemOpen
  \bibfield  {author} {\bibinfo {author} {\bibfnamefont {T.}~\bibnamefont
  {Kojo}}\ and\ \bibinfo {author} {\bibfnamefont {S.}~\bibnamefont {Itatani}},\
  }\bibfield  {title} {\bibinfo {title} {{Delineating neutral and charged
  mesons in magnetic fields}},\ }\href@noop {} {\  (\bibinfo {year} {2026})},\
  \Eprint {https://arxiv.org/abs/2604.15897} {arXiv:2604.15897 [hep-ph]}
  \BibitemShut {NoStop}%
\bibitem [{\citenamefont {Hattori}\ \emph {et~al.}(2016)\citenamefont
  {Hattori}, \citenamefont {Kojo},\ and\ \citenamefont {Su}}]{Hattori:2015aki}%
  \BibitemOpen
  \bibfield  {author} {\bibinfo {author} {\bibfnamefont {K.}~\bibnamefont
  {Hattori}}, \bibinfo {author} {\bibfnamefont {T.}~\bibnamefont {Kojo}},\ and\
  \bibinfo {author} {\bibfnamefont {N.}~\bibnamefont {Su}},\ }\bibfield
  {title} {\bibinfo {title} {{Mesons in strong magnetic fields: (I) General
  analyses}},\ }\href {https://doi.org/10.1016/j.nuclphysa.2016.03.016}
  {\bibfield  {journal} {\bibinfo  {journal} {Nucl. Phys. A}\ }\textbf
  {\bibinfo {volume} {951}},\ \bibinfo {pages} {1} (\bibinfo {year} {2016})},\
  \Eprint {https://arxiv.org/abs/1512.07361} {arXiv:1512.07361 [hep-ph]}
  \BibitemShut {NoStop}%
\end{thebibliography}%

\end{document}